\documentclass[a4paper,12pt]{article}
\usepackage[dvips]{graphicx}

\begin{document}
\begin{center}
{\large \bf Fractal proptrties of SDSS quasars}
\vspace{5ex}

I. K. Rozgacheva$^a$, A. A. Agapov$^b$
\vspace{2ex}

Moscow State Pedagogical University, Moscow, Russia
\vspace{1ex}

E-mail: $^a$rozgacheva@yandex.ru, $^b$agapov.87@mail.ru
\vspace{4ex}
\end{center}
\begin{abstract}

The distribution of SDSS quasars is described. The dependence of observational number of quasars within distances to an observer less than $r$ on distance $r$ for the flat Universe filled with cold dust is found to be the power law ${N(<r)\sim r^{2{,}71}}$ for the redshift range ${0,35<z<2,30}$. The quasar distribution on the celestial sphere is characterized by power laws as well: ${N(<\vartheta)\sim \left(\sin{\frac{\vartheta}2}\right)^{d_c}}$, where $d_c\approx 1{,}49\div 1{,}58$ for different redshift layers in the same range. These properties are evidences of fractality.
\end{abstract}
\vspace{1ex}

\vspace{3ex}

\begin{center}
1. Introduction
\end{center}

The aim of this work is investigation of the fractal properties of the Universe's large-scale structure by the quasars distribution. For this the seventh version of the SDSS catalogue \cite{SDSS} is used.

The analysis of available at present catalogues shows that galaxies clumps are characterized by a number of fractal properties \cite{SL}. Firstly, for every galaxy in a clump a number of neighbour galaxies $N$ in a volume with radius $r$ centered in this galaxy has closely power-law dependence on $r$:
$$
N(<r)\sim r^{d_c}.\eqno (1)
$$
The exponent (correlation dimension) has values in the range $d_c\approx 1{,}15\div 2{,}25$ for spatial scales $r\approx \left(1\div 10\right)h^{-1}Mpc$, where $h=\frac{H_0}{100\frac{km/s}{Mpc}}$, $H_0$ is the Hubble constant value at present [2].

The power law (1) is true for every clump, therefore it may be considered as a law of the large-scale structure.

The correlation dimension characterizes the degree of the galaxies clumping. The lower the correlation dimension the greater the galaxies distribution differs from the homogeneous distribution. In the late 20th century the correlation dimension was revealed according to small surveys containing not more than 10 thousand galaxies to increase with increasing radius approaching the value $d=3$ which corresponds to a homogeneous distribution. Therefore, the fractality of the large-scale is supposed to exist to distances $300h^{-1}Mpc$ (some observers estimates this distance as $1000h^{-1}Mpc$).

Secondly, the mean galaxy number density calculated according to different catalogues is found to depend on the catalogue sample volume. If the galaxy distribution is random (Poisson) the mean galaxy number density relates to the general totality and doesn't depend on the sample volume. The dependence of the mean number density on the catalogue may be related to the both the present catalogues contain a part of the real galaxies number in the selected sky area only (perhaps we see bright galaxies only) or the spatial galaxy distribution isn't random. For example, let galaxies form clumps characterized by the following properties. Every clump consists of $m_k$ galaxy pairs and the distance between the galaxies of the pair is $r_k$. Let clumps are related by the transformation
$$
m_k=pm_{k-1},\ r_k=qr_{k-1},\eqno (2)
$$
where $k=1,2,3,\dots$ and $p>1$, $q>1$ are constant numbers. Quantities $m_k$ and $r_k$ are evident to be a geometric progression:
$$
m_k=pm_{k-1}=m_0p^{k-1},\ r_k=qr_{k-1}=r_0q^{k-1},
$$
where $m_0$, $r_0$ are the normalization constants.

The total number of galaxies in the sample and the radius of the sample volume equal
$$
N=2\sum_{k=1}^{N/2}m_k=2m_0\sum_{k=1}^{N/2}p^{k-1}=2m_0\frac{p^{N/2}-1}{p-1}\approx 2m_0p^{\frac N2-1},
$$
$$
r_{N/2}=r_0q^{\frac N2-1}.
$$
Using these expressions one can deduce the expressions for galaxies number and their mean number density:
$$
N=2m_0\left(\frac{r_{N/2}}{r_0}\right)^{\frac{\ln{p}}{\ln{q}}}\sim \left(r_{N/2}\right)^{\frac{\ln{p}}{\ln{q}}},\eqno (3 a)
$$
$$
\langle n\rangle =\frac3{2\pi}m_0{r_0}^{-3}\left(\frac{r_{N/2}}{r_0}\right)^{\frac{\ln{p}}{\ln{q}}-3}\sim \left(r_{N/2}\right)^{\frac{\ln{p}}{\ln{q}}-3}.\eqno (3 b)
$$
These expressions shows that the galaxy mean number density depends on the sample volume and the observable correlation (1) is analogous to (3) if the constants $p$ and $q$ are chosen so that $d_c=\frac{\ln{p}}{\ln{q}}$. In this example the condition (2) is a condition of geometric similarity of clumps underlying the definition of a fractal set \cite{M,L,PGU}. The power law (1) is a consequence of the fractality condition (2), therefore power laws are called fractal laws.

The power type of the two-point correlation function is considered as the third feature of fractality. On scales from $0{,}1h^{-1}Mpc$ to  $10h^{-1}Mpc$ the observable galaxy two-point correlation function is a power law
$$
\xi (r)=A\left(\frac{r}{Mpc}\right)^{-\gamma },\eqno (4)
$$
where $15\le A\le 100$ and $1{,}6\le \gamma \le 2{,}2$ depending on a catalogue. The amplitude $A$ defines a correlation scale: $r_c=A^{1/\gamma }Mpc$. In clumps with high galaxy number density where $\xi >1$ the approximate equality $d_c=3-\gamma $ is true \cite{P}.

In section 2 of this paper the results of the new analysis of the quasar distribution according to the seventh version of the largest modern survey SDSS data \cite{SDSS} are adduced. According to the SDSS data there was an epoch of the high galaxy activity at redshift $0{,}35<z<2{,}30$ in the Universe evolution (see the histogram on fig. 1). Now receiving light from nucleus of the active galaxies – quasars we may consider them as markers of the space. The quasar distribution in this epoch proves to be characterized by the high correlation dimension value. A number of quasars within the observer’s light horizon grows with growing $r$ as the power law (1) with $d_c\approx 2{,}71$ in the high galaxy activity epoch. On the other hand, the correlation dimension of the galaxy distribution is not more than 2,25 in our neighbourhood within redshift $z<0,003$ \cite{CP}. This difference of the correlation dimension values indicates that clumping of the spatial galaxy distribution was lower in the past then in the present.

In section 2 the analysis of the quasar distribution on the celestial sphere for the SDSS sample is executed. The correlation dimension on the two-dimensional spherical surface turns out to equal approximately $d_c\approx 1{,}5$ on the average for quasar clumps.
\vspace{3ex}
\newpage

\begin{center}
2. The fractal properties of the observable quasar distribution
\end{center}

The seventh version of the SDSS catalogue containing 105761 quasars with redshifts $0{,}0645\le z\le 5{,}4608$ is used for the investigation of the quasar distribution. The catalogue was compiled by the wide-angle, narrow-angle and pencil beam surveys. For the purpose of the present work the area of the celestial sphere with the equatorial coordinates $7^h<\alpha <18^h$, $0^{\circ }<\delta <70^{\circ }$ covered by the wide-angle survey was chosen.

\begin{figure}[t]
\begin{center}
\includegraphics[scale=0.8]{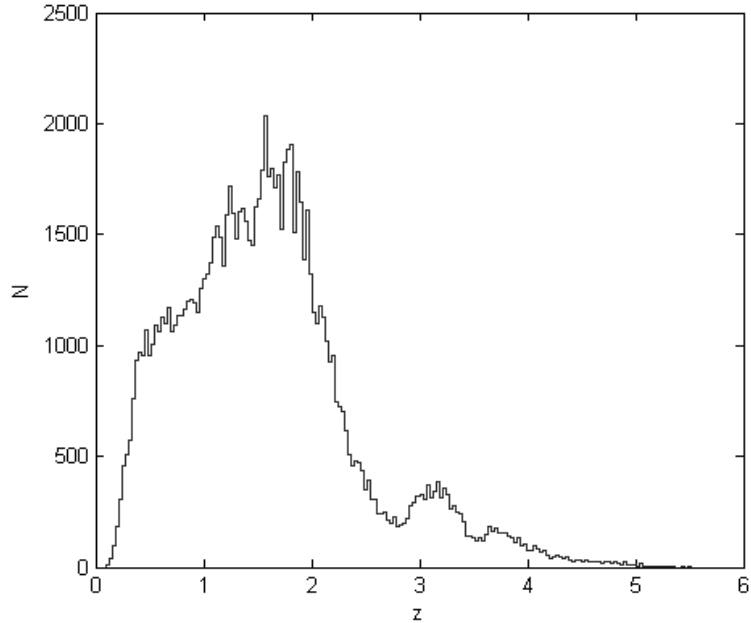}
\caption{The histogram of the redshift quasar distribution (SDSS catalogue)}
\end{center}
\end{figure}

The histogram of the quasar redshift distribution is adduced on fig.~1. The redshift range $0{,}35<z<2{,}30$ with the maximal quasar number is chosen for the further investigation. The $N$--$z$ relation is shown on fig.~2a, where $N$ is number of quasars with redshift from 0 to $z$.

The dependence of the observational number of quasars $N$ with distances to an observer less than $r$ on the distance $r$ is required to be found for estimating of the correlation dimension value.

There is the following $r$--$z$ correlation in the homogeneous and isotropic cosmological model with $\Lambda$-term and filled by cold dust (pressure $P=0$) and by relativistic gas not interacting with the dust \cite{ZN}:
$$
r(z)=\frac{c}{H_0}\int\limits_0^z\frac{dz}{\sqrt{\Omega_r\left(z+1\right)^4+\Omega_m\left(z+1\right)^3+\left(1-\Omega_m-\Omega_r-\Omega_\Lambda \right)\left(z+1\right)^2+\Omega_\Lambda}},
$$
where $\Omega_m$ and $\Omega_r$ are the non-dimensional density parameters of the dust and the relativistic gas respectively, $\Omega_\Lambda =\left(\frac{c}{H_0}\right)\frac{\Lambda}3$ ($\Lambda$ is the $\Lambda$-term in the Einstein's equations). The case of the flat Universe filled with the dust only where ${\Omega_r=\Omega_\Lambda =0}$ and $\Omega_m=1$ is considered here. Then
$$
r(z)=\frac{2c}{H_0}\left(1-\frac1{\sqrt{z+1}}\right).
$$
The dependence of $N$ on the non-dimensional distance $r'=\frac{r}{c/H_0}$ is presented on fig.~2b.

\begin{figure}[t]
\includegraphics[width=1\linewidth]{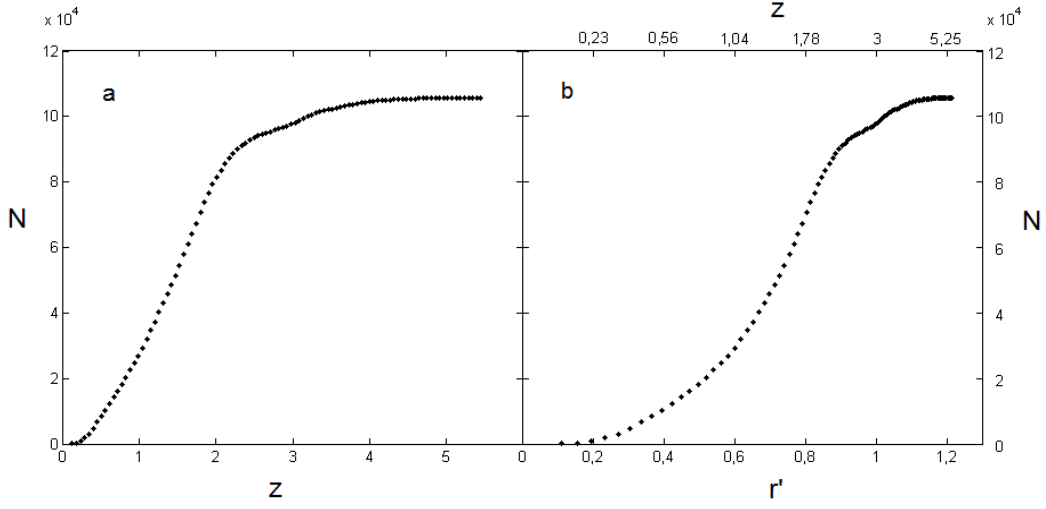}
\caption{a) $N$--$z$ relation; b) $N$--$r'$ relation}
\end{figure}

The logarithmic plot of the same relation for the redshift range $0{,}39<z<2{,}33$ and the least-squares line are shown on fig.~3. The angular coefficient of the line equals 2,71. The plot on fig.\ 2b consists of 100 dots, and 37 ones get into this redshift range. The correlation coefficient of the dots equals $R=0{,}999$. The critical value of the correlation coefficient equals $R_c=0{,}519$ when the number of degrees of freedom equals 35 and the significance level equals 0,001 \cite{BS}. As $R>R_c$ there is a significant statistical correlation between $N$ and $r'$ expressed by the power law
$$
N(<r)\sim r^{2{,}71}.
$$

The 12 redshift layers with thickness $\Delta z=0{,}03$ corresponding to the peaks of the histogram on fig.~1 in the same redshift range have been chosen for the studying of the quasar distribution on the celestial sphere. The quasar distribution on the celestial sphere for the redshift layer $1{,}23<z<1{,}26$ is shown on fig.~4 for example. There are regions of relative concentration and rarefaction in the distribution. Further, 10 quasars located approximately in the centers of concentration regions were chosen in every layer.

\begin{figure}[t]
  \begin{center}
  \includegraphics[scale=0.8]{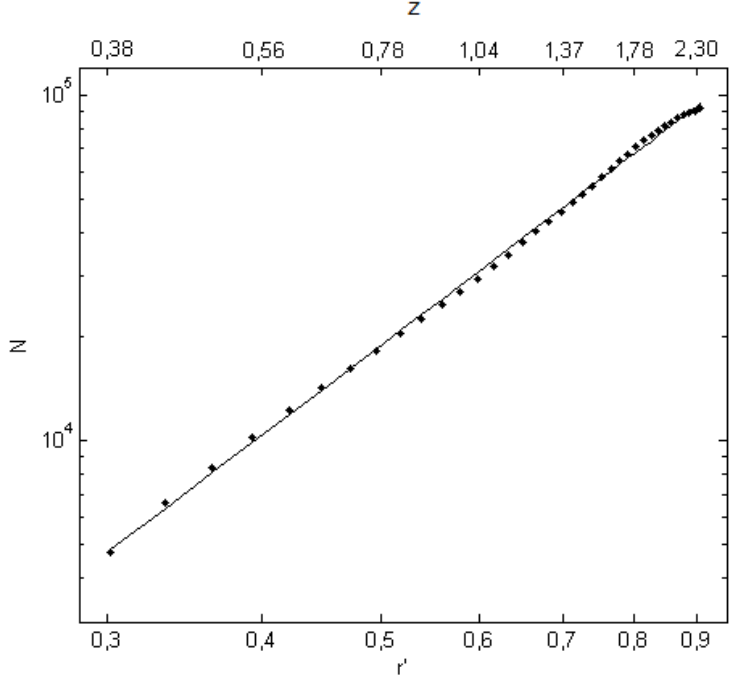}
  \caption{The logarithmic plot of $N$--$r'$ relation for the redshift range ${0{,}39<z<2{,}33}$}
  \end{center}
\end{figure}

\begin{figure}[t]
  \begin{center}
  \includegraphics[scale=0.8]{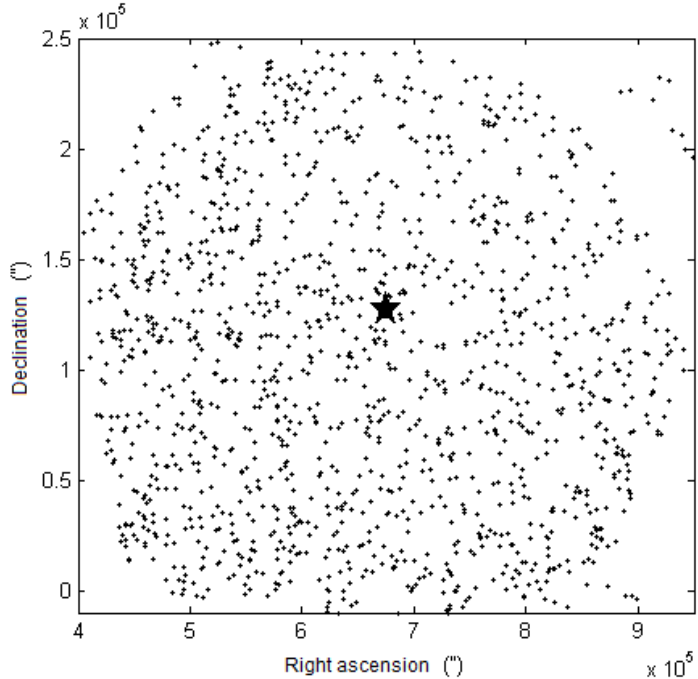}
  \caption{The quasar distribution on a part of the celestial sphere covered by the wide-angle survey in the redshift layer $1{,}23<z<1{,}26$ (the asterisk denotes the location of the quasar which the plot on fig.\ 5b concerns)}
  \end{center}
\end{figure}

Angular distance $\vartheta$ between every two quasars on the celestial sphere equals
$$
\vartheta = \arccos{\left(\cos{\delta_1}\cos{\delta_2}\cos{\left(\alpha_1-\alpha_2\right)}+\sin{\delta_1}\sin{\delta_2}\right)},
$$
where $\delta_1$, $\delta_2$, $\alpha_1$ and $\alpha_2$ are declinations and right ascensions of the first and the second quasar respectively.

The dependence of number of quasars $N$ with angular distances less than $\vartheta$ on $\sin{\left(\vartheta/2\right)}$ is required to be found for estimating of the correlation dimension value of the distribution. This operation has been carried out for each of 10 quasars in each of 12 redshift layers.

The mean angular distance between centers of clumps equals $30^{\circ}$ approximately. Therefore the maximal angular radius has been chosen $15^{\circ}$ in each case to decrease effects of getting the studied area by neighbour clumps. From 70 to 200 quasars get each area.

The examples of such dependences $N$ on $\sin{\left(\vartheta/2\right)}$ for four different redshift layers and for one clump in each of these layers are shown on fig.~5. The equatorial coordinates of clump's central quasar, a redshift range of a layer, an angular coefficient of a least-squares line and correlation coefficient value are presented in the caption.

The correlation coefficient $R$ values are not less than 0,950 for every clump in every redshift layer. But the correlation coefficient critical value equals $R_c=0{,}679$ when the number of degrees of freedom equals 18 and the significance level equals 0,001. As $R>R_c$ there is a significant statistical correlation between $N$ and $\sin{\left(\vartheta/2\right)}$ expressed by the power law
$$
N\left(<\vartheta\right)\sim \left(\sin{\frac{\vartheta}2}\right)^p.
$$

\begin{figure}[t!]
\includegraphics[width=1\linewidth]{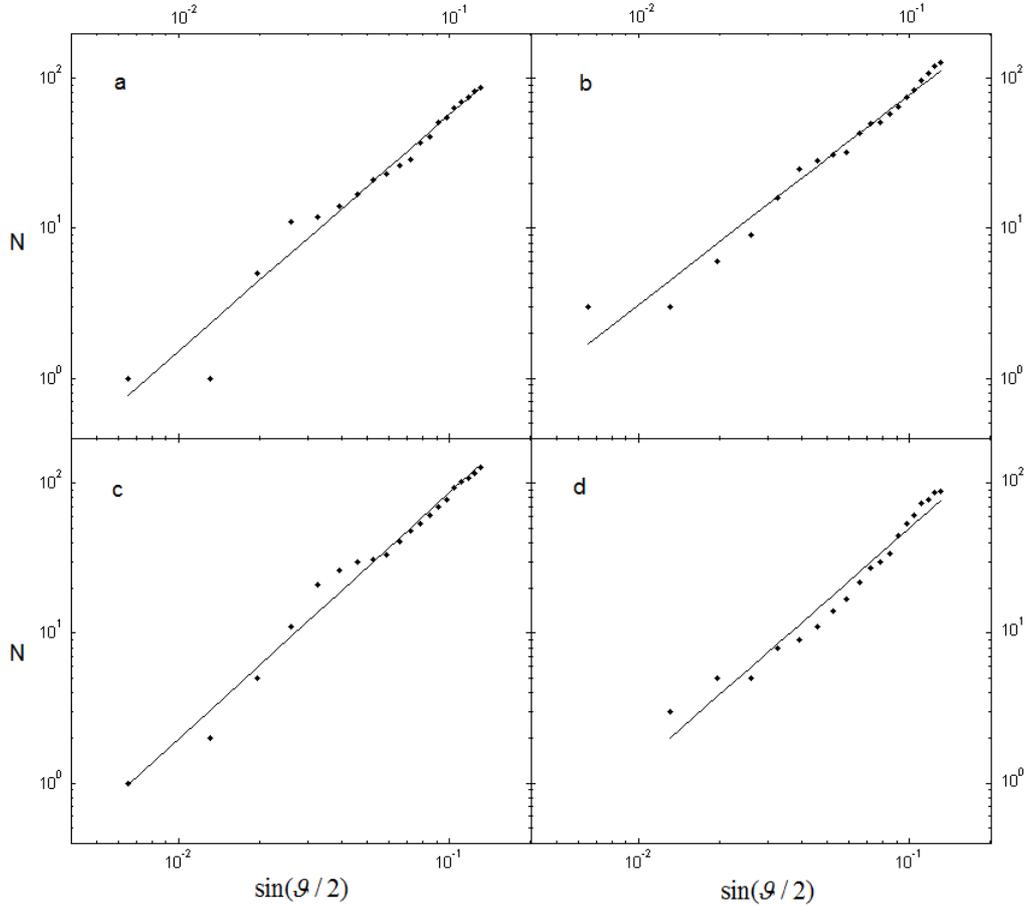}
\caption{$N$--$\sin{\vartheta/2}$ relation for different clumps in different redshift layers, $\alpha$ and $\delta$ are the equatorial coordinates of clump's central quasar:
a)~${\alpha=8^h48^m16^s}$,~${\delta=27^{\circ}45'56''}$,~${0{,}66<z<0{,}69}$,~${p=1{,}58}$,~${R=0{,}983}$;
b)~${\alpha=12^h30^m34^s}$,~${\delta=35^{\circ}10'5''}$,~${1{,}23<z<1{,}26}$,~${p=1{,}41}$,~${R=0{,}985}$;
c)~${\alpha=14^h49^m41^s}$,~${\delta=42^{\circ}22'54''}$,~${1{,}86<z<1{,}89}$,~${p=1{,}64}$,~${R=0{,}991}$;
d)~${\alpha=12^h25^m4^s}$,~${\delta=24^{\circ}21'44''}$,~${2{,}07<z<2{,}10}$,~${p=1{,}59}$,~${R=0{,}981}$.}
\end{figure}

Further, the correlation dimension value $d_c$ of each redshift layer has been estimated as the arithmetic mean of the angular coefficient $p$ values of the all 10 clumps in the layer. It takes the values from 1,49 to 1,56 for different layers.

The dependence of the correlation dimension on $z$ is presented on fig.~6. The correlation dimension grows with growing $z$ at the average as it should be if the clumping of quasars was lower in the past. The correlation coefficient value equals $R=0{,}250$. The correlation coefficient critical value equals $R_c=0{,}497$ when the number of degrees of freedom equals 10 and the significance level equals 0,1. Therefore, there is no significant statistical correlation between $d_c$ and $z$ and we cannot assert with confidence that the quasars clumping decreases with increasing $z$. Probably, a more complicated method of analysis of the quasar distribution on the celestial sphere, particularly a method of clumps separation, will allow to detect significant correlation.

\begin{figure}[t]
  \begin{center}
  \includegraphics[scale=0.8]{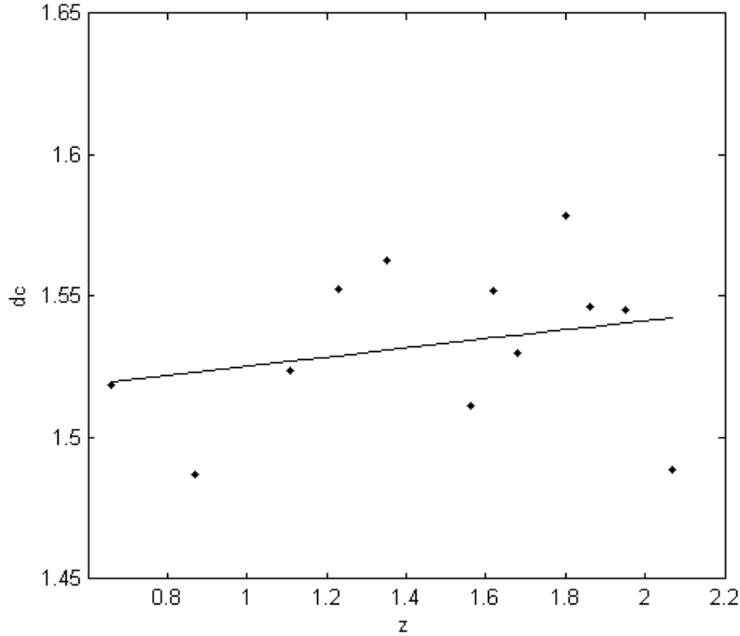}
  \caption{Dependence of the correlation dimension $d_c$ value on redshift $z$}
  \end{center}
\end{figure}
\vspace{3ex}

\begin{center}
3. Conclusion
\end{center}

The main results of the present work are following:

- the dependence of observable number of quasars with distances to an observer less than $r$ on the distance $r$ for the flat Universe filled with cold dust is revealed to be the power law ${N(<r)\sim r^{2{,}71}}$ at the high galaxy activity epoch, i.\ e. in the redshift range $0{,}35<z<2{,}30$;

- the quasar distribution on the celestial sphere is found to be of fractal character with correlation dimension from 1,49 to 1,58 for different $z$ layers.
\newpage

\begin{center}
{\bf Acknowledgments}
\end{center}

The authors are grateful to A.A. Borisov for his help in making of program texts for the data processing.

\end{document}